\documentclass[amsmath,amssymb,aps]{revtex4}

\usepackage{epsfig}
\usepackage{bbm}

\begin{document}

\title{Exact shock measures and steady-state selection in a driven
  diffusive system with two conserved densities} 

\author{A.\ R\'akos}
\author{G.\ M.\ Sch\"utz}

\affiliation{Institut f\"ur Festk\"orperforschung, Forschungszentrum J\"ulich - 
52425 J\"ulich, Germany}

\date{\today}

\begin{abstract}
We study driven 1d lattice gas models with two types of particles and nearest
neighbor hopping. We find the most general case when there is a shock
solution with a product measure which has a density-profile of a step
function for both densities. The position of the shock performs a biased
random walk. We calculate the microscopic hopping rates of the shock. We 
also construct the hydrodynamic limit of the model and solve the resulting 
hyperbolic system of conservation laws. In case of open boundaries the
selected steady state is given in terms of the boundary densities.  
\end{abstract}

\maketitle

\section{introduction}

On the macroscopic level driven diffusive systems are often described by
hydrodynamic equations for some relevant parameters (usually the particle
densities). These partial differential equations are generally
nonlinear and can develop shocks,   
i.e., discontinuities in the space dependence of the densities. These shocks
behave as collective excitations in systems with one conservation law:
they can be characterized by only one  
parameter, namely their position, and the propagation can be described by
single particle dynamics. 

Recently much attention was payed to the investigation of the microscopic
structure  and the microscopic dynamics of such shocks
\cite{Ferrari,DLS,Belitsky,KKrebs,Balazs}. Mostly the 
well-known asymmetric simple exclusion
process (ASEP) was studied: it was pointed out in \cite{KKrebs} (and for
infinite systems in \cite{Belitsky}) that for special tuning of densities and
microscopic hopping rates there exists a travelling shock with a
step-like density profile even on the microscopic scale, which behaves like a
one-particle excitation. However, little is known about the
microscopic structure of shocks in systems with
two conservation laws \cite{two-chanel,reflection,TothB,Fritz}, which 
recently have become a focus of attention  
(for a review see \cite{review}). 

In the present work we show that a shock measure with single-particle
dynamics can describe
also systems with two conserved densities, introduced in
section~\ref{sec:model}. The position of the     
shock performs a biased random walk just like in the ASEP
(section~\ref{sec:shockmeasure}). In section~\ref{hydro} we also study
the hydrodynamic limit of the model (under Eulerian scaling) which
shows a larger class of stable shock solutions. The hierarchical
structure of the hydrodynamic equations for this system allows us to
deduct the steady state selection in an open system connected to
particle reservoirs at its boundaries. 

\section{Two-species exclusion process}
\label{sec:model}

The model is defined on an open lattice with $L$ sites and two types of 
particles ($A$
and $B$). Each lattice site can either be occupied by a particle, which 
can be either $A$ or $B$, or can be vacant ($\varnothing$). 
Having two independent conserved quantity requires  
that only hopping processes are allowed. For simplicity we allow just
for nearest neighbour hoppings. These are the following: 
\begin{align}
  A\varnothing &\rightarrow \varnothing A & \varnothing A&\rightarrow A\varnothing  \cr
  B\varnothing &\rightarrow \varnothing B & \varnothing B&\rightarrow B\varnothing  \cr
  BA&\rightarrow AB & AB&\rightarrow BA 
\end{align}

A state of the model is defined through a probability measure
$P_\eta$ on the set of all configurations $\eta=(\eta_1,\eta_2, \dots,
\eta_L)$, $\eta_k\in\{A,B,\varnothing\}$.  For our purposes it is
convenient to use the Hamiltonian
formalism \cite{PTCP}  where
one assigns a basis vector $|\eta\rangle$ of the vector space
$(\mathbb{C}^3)^{\otimes L}$ to each configuration and the probability vector
is defined by $|P\rangle=\sum_\eta P_\eta |\eta\rangle$. It is
normalized such that $\langle s|P\rangle=1$ where
$\langle s|=\sum_\eta \langle\eta|$. The time dependence is now
described by the master equation 
\begin{equation}
  \label{master}
  \frac{d}{dt} |P(t)\rangle = -H |P(t)\rangle
\end{equation}
through a ``Hamiltonian'' $H$ which has matrix elements $H_{\eta,\eta'}$
the hopping rates between configurations $\eta,\eta'$.
Since we have only nearest neighbor exchange processes the
Hamiltonian in (\ref{master}) can be written as
\begin{equation}
  \label{H}
  H=h_1 + \sum_{k=1}^{L-1} h_{k,k+1}+ h_L,
\end{equation}
where $h_{k,k+1}$ acts nontrivially only on sites $k$ and $k+1$
(corresponding to hopping) while $h_1,h_L$ generates boundary processes
specified below. Let $W^{\theta_1
\theta_2}_{\theta_1' \theta_2'}$
($\theta_1,\theta_2,\theta_1',\theta_2'\in\{A,B,\varnothing\}$) be an operator
on $\mathbb{C}^3\otimes\mathbb{C}^3$ which acts on the basis vectors $|\eta_1,
\eta_2\rangle$ as
\begin{equation}
  \label{W}
  W^{\theta_1 \theta_2}_{\theta_1' \theta_2'}|\eta_1, \eta_2\rangle = 
  \delta_{\eta_1 \theta_1}\delta_{\eta_2 \theta_2} |\theta_1', \theta_2'\rangle.
\end{equation}
Then $h_{k,k+1}$ can be written as
\begin{equation}
  \label{h}
  h_{k,k+1}=\mathbbm{1}^{\otimes(k-1)}\otimes \left[\sum_{\theta_1 \theta_2 \theta_1' \theta_2'} 
  -\Gamma^{\theta_1 \theta_2}_{\theta_1' \theta_2'}
  \left(W^{\theta_1 \theta_2}_{\theta_1' \theta_2'} - 
  W^{\theta_1 \theta_2}_{\theta_1 \theta_2}\right)\right]
\otimes \mathbbm{1}^{\otimes(L-k-1)} 
\end{equation}
The diagonal term in (\ref{h}) stands for the conservation of
probability.  The model is defined through the rates $\Gamma^{\theta_1
\theta_2}_{\theta_1' \theta_2'}$ which describe the hopping process
$\theta_1 \theta_2 \rightarrow \theta_1' \theta_2'$. In our model the
nonzero rates are the following:
\begin{align*}
  \label{rates}
  \Gamma^{A\varnothing}_{\varnothing A}&=a_1 & \Gamma^{\varnothing
  A}_{A\varnothing}&=a_2 \cr 
  \Gamma^{B\varnothing}_{\varnothing B}&=b_1 & \Gamma^{\varnothing
  B}_{B\varnothing}&=b_2 \cr 
  \Gamma^{BA}_{AB}&=c_1 & \Gamma^{AB}_{BA}&=c_2 
\end{align*}
The boundary terms $h_1$ and $h_L$ in (\ref{H}) act only on the first
and $L$-th site respectively and choosing the basis  
\begin{equation}
  |A\rangle = 
  \left(\begin{array}{c}
    1 \\ 0 \\ 0
  \end{array}\right), \quad
  |B\rangle = 
  \left(\begin{array}{c}
    0 \\ 0 \\ 1
  \end{array}\right), \quad
  |\varnothing\rangle = 
  \left(\begin{array}{c}
    0 \\ 1 \\ 0
  \end{array}\right),
\end{equation}
they have the form
\begin{equation}
  \label{h_boundary}
  h_1=
  \left(\begin{array}{ccc}
    \beta_l^A+\gamma^-_l & -\alpha_l^A & -\gamma^+_l \\
    -\beta_l^A & \alpha_l^A + \alpha_l^B & -\beta_l^B \\
    -\gamma^-_l & -\alpha_l^B & \beta_l^B+\gamma^+_l
  \end{array}\right)\otimes \mathbbm{1}^{\otimes(L-1)}, \quad
  h_L=
   \mathbbm{1}.^{\otimes(L-1)}\otimes \left(\begin{array}{ccc}
    \beta_r^A+\gamma^-_r & -\alpha_r^A & -\gamma^+_r \\
    -\beta_r^A & \alpha_r^A + \alpha_r^B & -\beta_r^B \\
    -\gamma^-_r & -\alpha_r^B & \beta_r^B+\gamma^+_r
  \end{array}\right),
\end{equation}
where  $\alpha^{A(B)}$, $\beta^{A(B)}$ and $\gamma^{+(-)}$ are the
rates for the following processes:
\begin{align}
&\alpha^A:\ \varnothing \to A, & &\alpha^B:\ \varnothing \to B,\\
&\beta^A :\ A \to \varnothing, & &\beta^B :\ B \to \varnothing, \\
&\gamma^+:\ B \to A,           & &\gamma^-:\ A \to B, 
\end{align} 
and the indexes $l$ and $r$ indicate the left and right boundary respectively.

It is known that in the finite periodic system there is a family of
steady states which are product measures \cite{TothB, AHR}. 
We call $P_\eta$ a product measure if it has the form
\begin{equation}
  \label{productmeasure}
  P_\eta=P^{(1)}_{\eta_1}P^{(2)}_{\eta_2} \dots P^{(L)}_{\eta_L},
\end{equation}
where $P^{(k)}_{\eta_k}$ is a probability measure on $\{A,B,\varnothing\}$. This
means that the probability vector has the following direct product
form: 
\begin{equation}
  \label{productvector}
  |P\rangle =   
  \left(\begin{array}{c}
    p^A_1 \\ 1-p^A_1-p^B_1 \\ p^B_1
  \end{array}\right) 
  \otimes
  \left(\begin{array}{c}
    p^A_2 \\ 1-p^A_2-p^B_2 \\ p^B_2
  \end{array}\right) 
  \otimes
  \dots
  \otimes
  \left(\begin{array}{c}
    p^A_L \\ 1-p^A_L-p^B_L \\ p^B_L
  \end{array}\right),
\end{equation}
where $p^X_k$ is the probability of finding an ``X'' particle on site $k$.

Straightforward calculations show that for having a
stationary product measure with uniform densities
\begin{equation}
  \label{unidens}
  |P\rangle=
  \left(\begin{array}{c}
    \rho^A \\ 1-\rho^A-\rho^B \\ \rho^B
  \end{array}\right)^{\otimes L},
\end{equation}
 the rates have to satisfy the following condition  (\cite{TothB, AHR}):
\begin{equation}
  \label{cond_for_productmeasure}
  a_1-a_2-b_1+b_2+c_1-c_2=0.
\end{equation}
This is already a sufficient condition in case of an infinite chain or
periodic boundary conditions. For open boundaries there are some
additional restrictions for the boundary rates, for which we find:
\begin{align}
  \label{product_BC}
    (a_1-a_2)\rho^A(1-\rho^A)-(b_1-b_2)\rho^A\rho^B &= 
    \alpha_l^A(1-\rho^A-\rho^B)-(\beta_l^A+\gamma^-_l)\rho^A+\gamma^+_l\rho^B \\
    &=(\beta_r^A+\gamma^-_r)\rho^A- \alpha_r^A(1-\rho^A-\rho^B)-\gamma^+_r\rho^B \\
    (b_1-b_2)\rho^B(1-\rho^B)-(a_1-a_2)\rho^A\rho^B &= 
    \alpha_l^B(1-\rho^A-\rho^B)-(\beta_l^B+\gamma^+_l)\rho^B +\gamma^-_l\rho^A\\
    &=(\beta_r^B+\gamma^+_r)\rho^B-\alpha_r^B(1-\rho^A-\rho^B)-\gamma^-_r\rho^A.
\end{align}
The physical meaning of these equations is that the steady state currents of the
conserved densities have to be fitted at the boundaries. 
In the following we will assume that condition
(\ref{cond_for_productmeasure}) is fulfilled. For boundary rates as given
above we say that the system is in contact with a reservoir of
densities $\rho^{A,B}$.

\section{invariant shock measure}
\label{sec:shockmeasure}

Krebs et.\ al.\ in \cite{KKrebs} pointed out that in the ASEP
with open boundaries 
there is an invariant shock measure which is a product measure with a
step-like density profile; $\rho_{l(r)}$ on the left (right) of the
shock. This means that if $|P_k\rangle$ is a state
having the shock between the sites $k$ and $k+1$ then these states for
$k=0,1,2,\dots,L$ generate a subspace of the vector-space of states
which is invariant under time evolution and thus the many-particle
problem is reduced to a one-particle one. This property holds  if the
densities $\rho_{l,r}$ satisfy the condition
\begin{equation}
\frac{p}{q}=\frac{\rho_r(1-\rho_l)}{\rho_l(1-\rho_r)},
\end{equation}
where $p$ and $q$ are the hopping rates of the particles to the right resp.\ left.
The resulting one-particle dynamics
have a natural interpretation as a simple random walk of the shock
position.

In the following we search for the conditions under which there exists 
in our three-state model an invariant shock measure which is a product 
measure with a jump in the local particle density. To this end
let $|P_k\rangle$ be defined as
\begin{equation}
  \label{P_k}
  |P_k\rangle = 
  \left(\begin{array}{c}
    \rho^A_l \\ 1-\rho^A_l-\rho^B_l \\ \rho^B_l
  \end{array}\right)^{\otimes k} \otimes
  \left(\begin{array}{c}
    \rho^A_r \\ 1-\rho^A_r-\rho^B_r \\ \rho^B_r
  \end{array}\right)^{\otimes L-k}
\end{equation}
The probability vectors $|P_k\rangle$ define an $L+1$-dimensional
subspace of the vector space on which $H$ acts. 
Closure of this family of shock measures under the time evolution
generated by $H$ is equivalent to requiring
\begin{gather}
  \label{HP_k}
  H|P_k\rangle=
  -d_r|P_{k+1}\rangle-d_l|P_{k-1}\rangle+(d_r+d_l)|P_k\rangle 
\quad \text{for} \quad 1\leq k \leq L-1, \\
  H|P_0\rangle= -\bar{d_r} |P_1\rangle + \bar{d_r} |P_0\rangle \quad \text{and}
\label{br1}
 \\
  \quad H|P_L\rangle= -\bar{d_l} |P_{L-1}\rangle + \bar{d_l} |P_L\rangle.
\label{br2}
\end{gather} This is
easy to see because of having only nearest neighbour interactions
double hoppings of the shock position (e.g.\
$|P_k\rangle\rightarrow|P_{k+2}\rangle$) cannot occur. If this
condition can be met by an appropriate choice of model parameters the
quantities
$d_{l,r}$ and $\bar d_{l,r}$ are non-negative real
numbers representing the hopping rates of the shock in the bulk
($d_{l,r}$) and at the boundaries ($\bar d_{l,r}$).
 
To investigate whether (\ref{HP_k}-\ref{br2}) can be satisfied it is 
better to rewrite the Hamiltonian in the form 
\begin{equation}
  \label{Hami}
  H=h_1+(a_1-a_2)n_1^A+(b_1-b_2)n_1^B +\sum_{k=1}^{L-1}\tilde h_{k,k+1} + 
  h_L-(a_1-a_2)n_L^A-(b_1-b_2)n_L^B,
\end{equation}
where
\begin{equation}
  \label{htilde}
  \tilde h_{k,k+1}=h_{k,k+1}-(a_1-a_2)(n_k^A-n_{k+1}^A)-(b_1-b_2)(n_k^B-n_{k+1}^B).
\end{equation}
Here and in (\ref{Hami}) $n_k^A$ and $n_k^B$ are the particle number
operators for the A and B particles on site $k$. 

As a result $\tilde h_{k,k+1}|P_l\rangle=0$ for $k\neq l$. Therefore
one has
\begin{equation}
  \label{HP_k'}
  H|P_k\rangle=
\begin{cases}
\tilde h_{k,k+1}|P_k\rangle 
  + \left(h_1+h_L+(a_1-a_2)(n_1^A-n_L^A)+
    (b_1-b_2)(n_1^B-n_L^B)\right)|P_k\rangle & \text{if } 1\leq k\leq L-1 \\
\left(h_1+h_L+(a_1-a_2)(n_1^A-n_L^A)+
    (b_1-b_2)(n_1^B-n_L^B)\right)|P_k\rangle & \text{if } k=0,L
\end{cases}
\end{equation}
At the boundaries one has to satisfy (\ref{product_BC})
with the slight modification that on the left (right) boundary we
write $\rho^{A,B}_l$ ($\rho^{A,B}_r$) instead of $\rho^{A,B}$: 
\begin{align}
  \label{boundary_a}
   \alpha_l^A(1-\rho^A_l-\rho^B_l)-(\beta_l^A+\gamma^-_l)\rho^A_l +\gamma_l^+\rho_l^B&=
  (a_1-a_2)\rho^A_l(1-\rho^A_l)-(b_1-b_2)\rho^A_l\rho^B_l \cr
  (\beta_r^A+\gamma_r^-)\rho^A_r-\gamma_r^+\rho^B_r- \alpha_r^A(1-\rho^A_r-\rho^B_r) &= 
  (a_1-a_2)\rho^A_r(1-\rho^A_r)-(b_1-b_2)\rho^A_r\rho^B_r,
\end{align}
and the same for the B particles:
\begin{align}
  \label{boundary_b}
     \alpha_l^B(1-\rho^A_l-\rho^B_l)-(\beta_l^B+\gamma^+_l)\rho^B_l +\gamma_l^-\rho_l^A&=
  (b_1-b_2)\rho^B_l(1-\rho^B_l)-(a_1-a_2)\rho^A_l\rho^B_l \cr
  (\beta_r^B+\gamma^+_r)\rho^B_r-\gamma^-_r\rho^A_r- \alpha_r^B(1-\rho^A_r-\rho^B_r) &= 
  (b_1-b_2)\rho^B_r(1-\rho^B_r)-(a_1-a_2)\rho^A_r\rho^B_r.
\end{align}
This fixes a manifold of boundary parameters required to satisfy
(\ref{HP_k}-\ref{br2}). The physical interpretation is a connection
to boundary reservoirs with different left and right densities
respectively.
Requiring (\ref{boundary_a}-\ref{boundary_b}) 
for the boundary operators $h_1$ and $h_L$ simplifies (\ref{HP_k'})
and leads to 
\begin{equation}
  \label{second}
  H|P_k\rangle = \tilde h_{k,k+1}|P_k\rangle +
  \left((a_1-a_2)(\rho^A_l-\rho^A_r)+(b_1-b_2)(\rho^B_l-\rho^B_r)\right)|P_k\rangle
  \quad \text{for } 1\leq k\leq L-1.
\end{equation}
Thus satisfying (\ref{HP_k}) is equivalent to requiring
\begin{multline}
  \tilde h_{k,k+1}|P_k\rangle = -d_r|P_{k+1}\rangle - 
  d_l|P_{k-1}\rangle \\
  +\left(d_r+d_l-(a_1-a_2)(\rho^A_l-\rho^A_r)-
  (b_1-b_2)(\rho^B_l-\rho^B_r)\right)|P_k\rangle, 
  \quad \text{for } 1\leq k \leq L-1
\end{multline}
Straightforward but lengthy calculation shows that $\tilde
h_{k,k+1}|P_k\rangle$ has this form for $1\leq k\leq L-1$
if and only if
\begin{gather} 
    \label{criterion1}
    a_1=b_1=:p, \quad
    a_2=b_2=:q, \\
    \label{criterion2}
    \frac{p}{q} = \frac{\rho_r(1-\rho_l)}{\rho_l(1-\rho_r)} \\
    \label{criterion3}
    \frac{\rho^A_l}{\rho^B_l} = \frac{\rho^A_r}{\rho^B_r} =: r,
\end{gather}
where $\rho_{l(r)}=\rho^A_{l(r)}+\rho^B_{l(r)}$ is the total density of
particles on the left (right) of the shock. We note here that
(\ref{criterion1}) together with (\ref{cond_for_productmeasure}) gives
also $c_1=c_2$. This means that the A and B particles behave the same
way: the hopping rates to the left and to the right are $p$ and
$q$ for both and also the exchange rates $c_1$ and $c_2$ are the
same in both directions. Equations (\ref{br1},\ref{br2}) give extra
conditions for the boundary rates, namely: 
\begin{equation}
\alpha_l^A =r \alpha_l^B \quad \text{and} \quad \alpha_r^A =r \alpha_r^B
\end{equation}
This together with (\ref{boundary_a}, \ref{boundary_b}) give
conditions also for the boundary rates $\beta_{l,r}^{A,B}$ and
$\gamma_{l,r}^{+-}$.

The rates $d_r$ and $d_l$ which describe the diffusion of
the shock position to the right and to the left are
\begin{equation}
  \label{gamma}
  d_r=\frac{\rho_r}{\rho_l} q = \frac{1-\rho_r}{1-\rho_l} p, 
  \quad d_l=\frac{\rho_l}{\rho_r} p = \frac{1-\rho_l}{1-\rho_r} q.
\end{equation}
In addition we find the boundary rates
\begin{gather}
  \bar{d_r}=\frac{\alpha_l}{\rho_l}-\rho_r(p-q)=(p-q)(1-\rho_r)+\frac{r\beta_l^A
  + \beta_l^B}{(r+1)(1-\rho_l)}, \\
  \bar{d_l}=\frac{\alpha_r}{\rho_r}+\rho_l(p-q)=-(p-q)(1-\rho_l)+\frac{r\beta_r^A
  + \beta_r^B}{(r+1)(1-\rho_r)},
\end{gather}
where $\alpha_{l(r)}=\alpha^A_{l(r)}+\alpha^B_{l(r)}$.

Using (\ref{criterion2}) the bulk shock hopping rates
(\ref{gamma}) become 
\begin{equation}
\label{asepshock}
d_{l(r)}=(p-q)\frac{\rho_{l(r)}(1-\rho_{l(r)})}{\rho_r-\rho_l}
\end{equation}
as in the ASEP \cite{KKrebs,Belitsky}. Since by 
identifying A and B particles one arrives at the one-species ASEP, it
is not surprising that  
(\ref{criterion2}) and (\ref{asepshock}) are in full agreement with
the corresponding formulas of the ASEP.
The novelty is that {\em a.)} the product
measure with step-like density profiles satisfying (\ref{criterion3})
remains an invariant shock measure even if one distinguishes between A and
B particles {\em b.)} there is no other possibility for having such an
invariant shock measure (up to relabelling the states $A$, $B$ and 
$\varnothing$).
Notice that this solution describes a {\em single} shock whereas
generically a system with two conservation is expected to have two
stationary shocks \cite{two-chanel,Fritz}.

\section{Hydrodynamic limit}
\label{hydro}

\subsection{Derivation of the partial differential equation}

In the previous sections the microscopic dynamics of the two-species
ASEP 
was studied. It was found that invariant product measures with step-like 
density 
profiles are possible if criterion (\ref{criterion1}) for the rates
and criteria (\ref{criterion2}), (\ref{criterion3}) for the
densities are fulfilled.
It is also of interest to drop the conditions (\ref{criterion2}), 
(\ref{criterion3}) and see how this model behaves on the 
macroscopic scale, i.e., when the space and time is rescaled as 
$t_\text{mac}=t_\text{mic} a$, $x=ka$ and the lattice spacing $a\to 0$ 
(Eulerian scaling). In this subsection we restrict the discussion to the
case of infinite chains (or torus geometry). Finite systems with open
boundaries are considered in the last subsection. 

Performing the Euler scaling one can easily construct the naive hydrodynamic 
equations of a model having two conserved quantities ($v$ and $u$) 
and stationary product measure. These 
describe the macroscopic time-evolution of the conserved quantities: 
\begin{align}
  \label{conslaw1}
  \partial_t v + \partial_x j_v(v,u)& = 0 \\
  \label{conslaw2}
  \partial_t u + \partial_x j_u(v,u)& = 0,
\end{align}
where $j_v(v,u)$ and $j_u(v,u)$ are the currents of $v$ and $u$ assuming homogeneous 
product measure. We have to note 
here that it is far from trivial that these set of 
conservation laws describe correctly the macroscopic dynamics of the 
model. However,
 a mathematically rigorous
proof is available that in all systems having nearest neighbor
dynamics with two conserved  
densities ($v$ and $u$) and stationary product measure, the time evolution 
of the densities under Eulerian scaling is
described by the equations (\ref{conslaw1},\ref{conslaw2}) until the 
occurrence of the first shock \cite{TothB}. For closely related
results on two-species systems, see \cite{Gross}, for an
apparent failure of the hydrodynamic description, see \cite{Hanney}
and earlier work on condensation in two-species systems 
\cite{review,Evan00,Muka00}.

For our model it is useful to introduce the new set of conserved
quantities $(v,u)$ instead of $(\rho^A,\rho^B)$:
\begin{align}
  v& = 1-\rho^A-\rho^B \quad (\text{the density of vacancies}) \\
  u& = \rho^A-\rho^B.
\end{align}
The associated currents are:
\begin{align}
  j_v& = -(p-q)v(1-v) \\
  j_u& = (p-q)vu.
\end{align}
Plugging these into (\ref{conslaw1},\ref{conslaw2}) one arrives to the hydrodynamic 
equations of the model:
\begin{align}
  \label{HD1}
  \partial_t v - (p-q)\partial_x(v(1-v))& = 0 \\
  \label{HD2}
  \partial_t u + (p-q)\partial_x (vu)& = 0.
\end{align}
In the special case when $p=q$ the currents are zero and the solution of 
(\ref{conslaw1},\ref{conslaw2}) becomes trivial. In this case the behaviour of 
the microscopic model is diffusive so one has to perform diffusive
scaling ($t_\text{mac}=t_\text{mic} a^2$, $x=ka$, $a\to 0$) to see 
the nontrivial macroscopic behaviour.

It is to be noted that the generic set of PDEs describing systems with
three local states and satisfying (\ref{cond_for_productmeasure}) is
the so-called Leroux's system \cite{TothB,Fritz}:
\begin{align}
  \label{Leroux1}
  \partial_t \sigma + \partial_x(\sigma\tau)& = 0 \\
  \label{Leroux2}
  \partial_t \tau + \partial_x (\sigma+\tau^2)& = 0.
\end{align} 
It means that there are always such linear combinations ($\sigma,\tau$) of the conserved
quantities (shifted by an irrelevant constant) which satisfy
(\ref{Leroux1}-\ref{Leroux2}) \cite{TothB}.   
However, in our case, when $a_1=b_1$ and
$a_2=b_2$, $\sigma$ and $\tau$ turn out to be linearly dependent and
equations (\ref{Leroux1}) and (\ref{Leroux2}) are equivalent (and
correspond to (\ref{HD1})). Therefore our model has to be investigated
separately.

Note that the equation  (\ref{HD1}) for $v$ is decoupled from $u$
and takes the form of the well-known  
Burgers equation, which can be solved exactly
\cite{Burgers}. Introducing 
\begin{equation}
\label{phi}
\phi=(\rho^A-\rho^B)/(\rho^A+\rho^B)=u/(1-v)
\end{equation} 
one
gets the linear wave equation for $\phi$:
\begin{equation}
\partial_t\phi+(p-q)v\partial_x\phi=0
\end{equation}
meaning that $\phi$ is constant along the curves $x(t)$ satisfying the
ordinary differential equation: 
\begin{equation}
\label{ODE}
\frac{dx}{dt} = v(x,t),
\end{equation}
where $v(x,t)$ is the solution of (\ref{HD1}). The physical meaning of 
$x(t)$ is that this is the expected path of a single tagged particle
(either $A$ or $B$).

\subsection{Development of an initial sharp interface}

Suppose that at $t=0$ there is a sharp interface separating domains characterised by 
$v_{l(r)}$ and $u_{l(r)}$ on the left (right) respectively. We assume that
$v_l>v_r$ in order to guarantee the stability of the interface 
\cite{two-chanel,PTCP}. No further conditions on the values of
$u,v$ are imposed.

\begin{figure}
\centerline{\epsfig{file=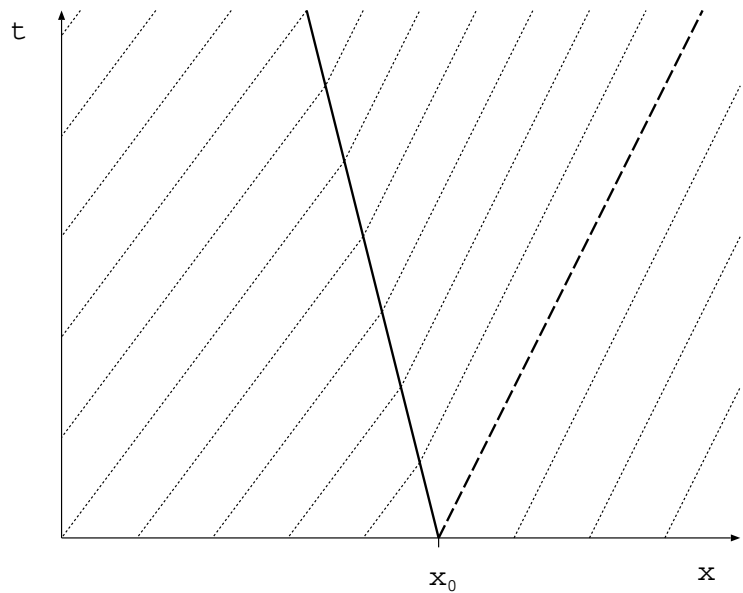,width=10truecm}}
\caption{The $x(t)$ curves of equation (\ref{ODE}) for the case when
  $v(x,t)$ is a shock-solution of the Burgers equation
  (\ref{HD1}). The position of this shock (starting from $x_0$) is
  marked by the thick solid line. $u$ also changes on the dashed line
  indicating another shock in the system, but this is not
  microscopically sharp.}\label{fig:shock} 
\end{figure}

The time and space dependence of $v$ is known since it is described by the inviscid 
Burgers equation (\ref{HD1}): 
the step-like front (having $v_{l(r)}$ on the left (right) of it) 
is travelling with velocity 
\begin{equation}
\label{shockvel1}
V_1=(p-q)(1-v_l-v_r).
\end{equation} 
The $x(t)$ curves
defined by equation (\ref{ODE}) are then  
broken lines as in figure \ref{fig:shock}. One can see that $\phi$
changes while crossing the bold dashed line indicating that there  
is another shock in the system travelling with larger velocity (without loss of generality we assume that $p>q$)
\begin{equation}
\label{shockvel2}
V_2=(p-q)v_l>V_1.
\end{equation}

Generally a single sharp interface at $t=0$ forms two shocks 
(or rarefaction waves) at $t>0$ in systems having two conservation
laws \cite{two-chanel}. For the complete analysis one has to study
the eigenvalues and eigenvectors of the Jacobian 
\begin{equation}
\frac{\partial (j_u,j_v)}{\partial(u,v)}=(p-q)\left( 
\begin{array}{cc}
v & u \\
0 & 2v-1 
\end{array}
\right)
\end{equation}
which give the characteristic velocities of the different types of
density fluctuations \cite{two-chanel}. 
In this specific system these are: 
\begin{align}
\lambda_1&=(p-q)(2v-1) & e_1&=\left(\begin{array}{cc} u \\ v-1 
\end{array}\right) \\
\label{secondmode}
\lambda_2&=(p-q)v  & e_2&=\left(\begin{array}{cc} 1 \\ 0 
\end{array}\right)
\end{align}
The two corresponding types of shocks have either constant $\phi$
across the discontinuity (type 1, occurring along
the solid bold line in Fig. 1) or constant $v$ (type 2, occurring along
the broken line in Fig. 1). A shock (microscopically
sharp interface) satisfying either of these relations
initially does not split, whereas a generic shock splits into two
shocks of type 1 and type 2 respectively.
We remark that condition (\ref{criterion3})
translates into $\phi_1=\phi_2$ for the quantity $\phi$. Hence the
shock studied in the previous section is of type 1.
Notice also that
in the second mode $u$ is changed but $v$ is not. Walking along
these lines in the $(u,v)$ space $\lambda_1$ remains unchanged. This
means that if there is a discontinuity in $u$ (with constant $v$)
then the fluctuations in $u$ are not driven towards the "shock" but
they travel with the same mean velocity resulting in an unstable shock
on the microscopic scale.   
However, since in the original lattice model the width of the step is
expected to scale 
with $\sqrt{t}$ due to the diffusive dynamics of $A$ and $B$
particles the interface remains sharp under Eulerian scaling.
 
\begin{figure}
\centerline{$t=0$ \epsfig{file=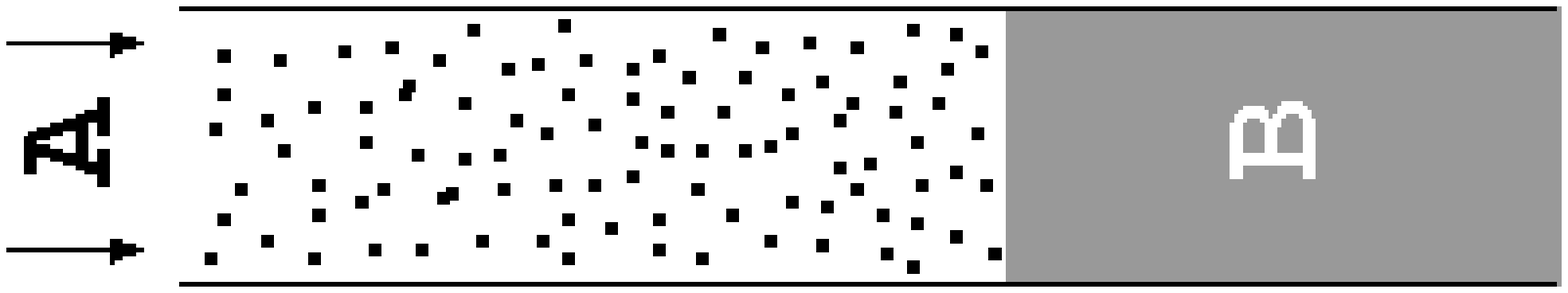,width=7truecm,angle=270}
\hspace{2cm}
$t>0$ \epsfig{file=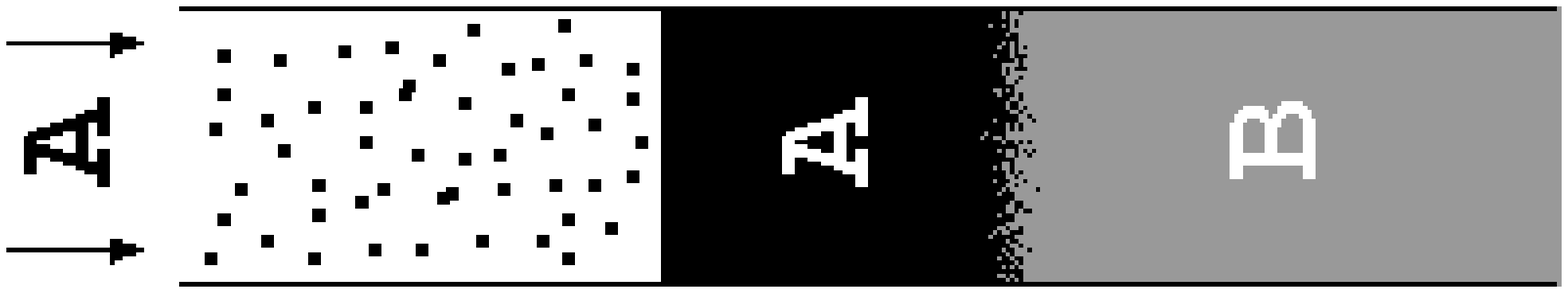,width=7truecm,angle=270}}
\caption{The time development from a special initial condition: $A$
  particles fall down (e.g.\ due to gravity) and enter a 
  medium where they have no weight, hence they penetrate diffusively
  but pile up ballistically resulting in two interfaces. The width of
  the domain wall between the $A$ (black) and $B$ (gray) particles
  scales with $\sqrt{t}$ while the other domain wall remains
  microscopically sharp (however, on a diffusive scale one could see
  the fluctuations of the position of this sharp interface).}
\label{fig:shock3}
\end{figure}

The physical significance of two types of shocks can be easily 
demonstrated considering the
following example: at 
$t=0$ the system is     
partially occupied with only $A$ particles on the left of $x_0$ and
fully occupied by $B$ particles on the right. This setting 
models a situation
where particles ($A$) fall down (e.g.\ due to gravity) and enter a 
medium ($B$) where they have no weight. Hence they penetrate diffusively
(with vanishing macroscopic flux), but pile up ballistically due to
the constant incoming macroscopic flux. The result are two interfaces 
emerging (see figure
\ref{fig:shock3}). The second (lower) interface has velocity 0 since
$v_1=v_2=0$, see (\ref{shockvel2}).
Although the particles diffuse and start mixing on the lattice scale
at the $A-B$ interface, this remains invisible on the Euler scale. 

The other interface (type 1) separating domains with different densities 
of $A$ particles corresponds to the usual
shocks which are known from the study of the Burgers
equation. Here both $u$ and $v$ are changed (but $\phi$
remains constant). 

\begin{figure}
\centerline{\epsfig{file=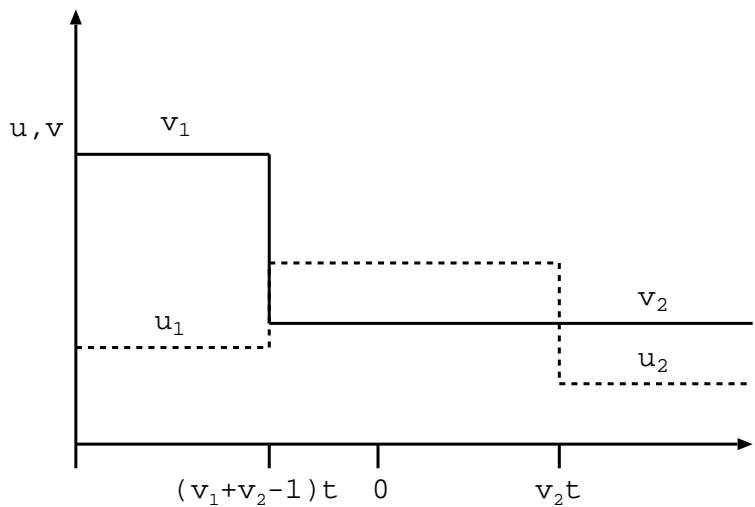,width=10truecm}}
\caption{Development of the densities $v$ and $u$ under Eulerian
  scaling starting from an initial state with step-like density
  profiles: $v_1, u_1$ on the left and $v_2, u_2$ on the right of the
  origin. The first interface (on the left) is a shock coming from the
  Burgers-equation and corresponds to the thick solid line on figure
  \ref{fig:shock}. The second one (on the right) is not
  microscopically sharp. On figure \ref{fig:shock} it is the dashed
  line.} \label{fig:shock2} 
\end{figure}

On figure \ref{fig:shock2} we show a more general example for the time
development 
of the system starting from an initial state having a step in both
densities at the origin.  The second mode also can develop rarefaction 
waves if the
density on the left is lower than on the right in the initial
state. We note that the shock coming from the second mode always stays
to the right of the shock/rarefaction wave coming from the first mode
since it is faster (\ref{shockvel2}). 
This is by the relation $v\geq 2v-1$ (rarefaction
waves) and $v_{1,2}\geq v_1+v_2-1$ (shocks).   

\subsection{Steady state selection in the open system}
\label{subsec:steady_selection}

In a finite system with open boundaries one is left with the question
of steady state selection. In the infinite system all the states with
constant density profiles are stationary, however in the finite system with
boundaries acting as particle reservoirs the bulk densities depend
non-trivially on the boundary densities. Setting $\partial_t=0$ in
(\ref{conslaw1})(\ref{conslaw2}) one arrives at the solution that $u$
and $v$ are constant, so in general the densities cannot fit both
boundaries. This discrepancy is resolved by the appearance of shocks
in the original lattice model leading to discontinuities at either (or
both) boundary in the hydrodynamic limit.    

In case of one conserved density the density-current relation already
determines the phase diagram in terms of the boundary densities
\cite{Kol, min} but in case of more conservation laws the question
turns out to be much more intricate and no general rule is known to
apply \cite{reflection}. However, in our model it is possible to
determine the resulting 
steady state for given boundary densities.   

Since the dynamics of $v$ is independent of $u$ and follows the usual
ASEP dynamics the profile can be deduced from the ASEP phase
diagram \cite{Liggett,DEHP,SD}. 
Our task is then to determine the bulk value of $u$ in terms of the
boundary densities
$u_\text{left},u_\text{right},v_\text{left},v_\text{right}$. 

Since all
the fluctuations in $u$ travel with velocity $v\geq 0$ 
one would think that  $u_\text{bulk}$ is only
determined by the left boundary if $v\neq0$. However, this is not the
case because a 
discontinuity in $v$ can be localised at this boundary which also induce a
step in $u$. The previously introduced $\phi$, however, does not
change at this discontinuity which suggests that the bulk value of $\phi$ is
given by the left boundary ($\phi_\text{bulk}=\phi_\text{left}$). 
In order to verify this heuristic reasoning we make use of a
technique to determine the steady state which is widely used
for systems with a single conservation law. One introduces
a viscosity term in (\ref{conslaw1},\ref{conslaw2}) proportional to
$\nu$ which contain a second derivative of the densities leading to
second order equations for the stationary profiles. Now the profiles
can fit both boundaries for any value of $\nu$ and after performing
limit $\nu\to 0$ one can get the selected density.  

In case of one single conservation law this technique is quite robust
in the sense that the resulting bulk density is essentially
independent of the specific choice of the viscosity term (as long as
it contains a second order derivative). For more conservation laws 
this method is mathematically not well-studied and one
has to be more careful with the choice of viscosity term.
For systems with a stationary product measure there is a natural choice 
for these terms introduced in \cite{two-chanel}. Namely, when one derives 
the hydrodynamic equations from the
lattice continuity equations one uses a Taylor expansion of the
current (as a function of the 
densities) in the lattice constant. Keeping the second order terms leads 
to a unique viscosity
term which is proportional to the lattice constant $a$.  

\begin{figure}
\centerline{\epsfig{file=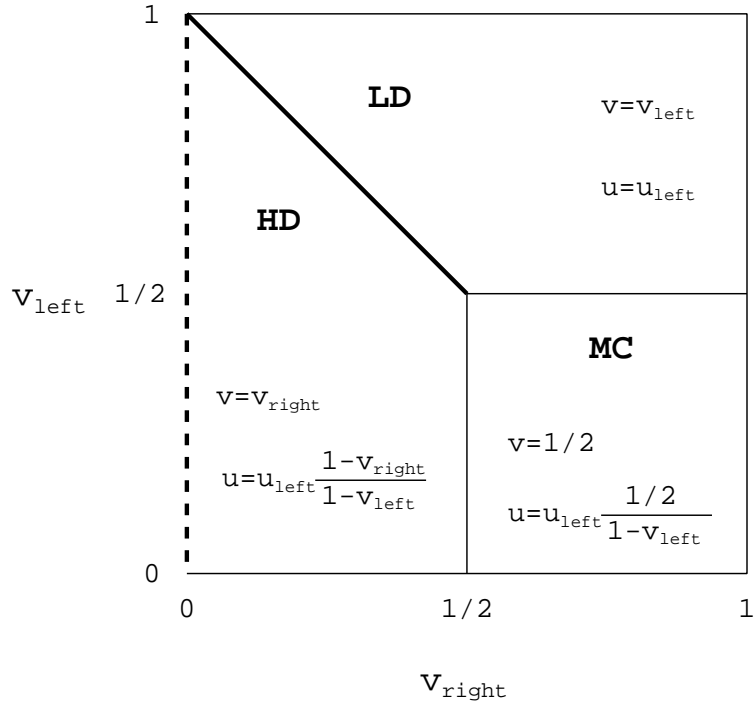,width=10truecm}}
\caption{The phase diagram of the open system. Since the vacancies
  follow the ASEP dynamics the boundary values of $v$ determine the
  phase according to the ASEP phase diagram. In addition equation
  (\ref{uprofile}) determines the stationary value of $u$ in all phases
  except on the dashed line where one has symmetric diffusion of $A$
  and $B$ resulting in linear profiles for $u$ connecting
  $u_\text{left}/(1-v_\text{left})$ and $u_\text{right}$.} 
\label{fig:phase_diagram} 
\end{figure}

In appendix \ref{visc} we show that using this technique for our model
the above conjecture, that
$\phi_\text{bulk}=\phi_\text{left}$, is confirmed if
$v_\text{right}\neq 0$, which implies 
\begin{equation}
\label{uprofile}
u=u_\text{left}\frac{1-v}{1-v_\text{left}}\quad\text{if }v_\text{right}\neq 0.
\end{equation}
Summarizing the results (see also figure
\ref{fig:phase_diagram}) we find
\begin{align}
v_\text{bulk}&=
\begin{cases} 
v_\text{left} & \text{if } v_\text{left}\geq 1/2 \text{ and }
v_\text{right} > 1-v_\text{left} \\ 
v_\text{right} & \text{if } v_\text{right}\leq 1/2 \text{ and }
v_\text{right} < 1-v_\text{left} \\ 
1/2 &  \text{if } v_\text{left}\leq 1/2 \text{ and } v_\text{right}\geq 1/2
\end{cases}, \\
u_\text{bulk}&=u_\text{left} \frac{1-v_\text{bulk}}{1-v_\text{left}}=
\begin{cases}
u_\text{left} & \text{if } v_\text{left}\geq 1/2 \text{ and }
v_\text{right} > 1-v_\text{left} \\
u_\text{left}\frac{1-v_\text{right}}{1-v_\text{left}} & \text{if }
0<v_\text{right}\leq 1/2 \text{ and } v_\text{right} < 1-v_\text{left}
\\
u_\text{left}\frac{1/2}{1-v_\text{left}} & \text{if }
v_\text{left}\leq 1/2 \text{ and } v_\text{right}\geq 1/2 
\end{cases}.
\end{align}

On the coexistence line 
($0<v_\text{right}=1-v_\text{left}<1/2$) the stationary $v$-profile
is linear as known from the ASEP \cite{SD,DEHP}. 
This induces also a linear profile for $u$ according to (\ref{uprofile}).

If $v_\text{right}=0$ then there are no vacancies in the system
leading to symmetric diffusion of $A$ and $B$ particles. This results
in linear profiles for $u$ connecting
$u_\text{left}/(1-v_\text{left})$, which is the value of $u$ after the
discontinuity located at the left boundary, and
$u_\text{right}$ as expected from the symmetric exclusion process (SEP) 
\cite{Spoh83} and
confirmed in appendix \ref{visc}.

\section{Conclusion}

We have introduced an asymmetric exclusion process with two conserved
species of particles. 
Its study is part of the program to understand
the emergence of macroscopic behaviour from microscopic models
of nonequilibrium diffusive interacting particles
and to elucidate the significance of boundary conditions in driven
diffusive systems. Under heuristic Eulerian scaling we obtain a system of
PDE's which can be analysed analytically.
As one would have expected from the nonlinear current-density relation
the system may produce shock discontinuities on the macroscopic scale.
There are two types of shocks (as expected in systems with two 
conservation
laws) for which we calculate the respective shock velocities and
intermediate densities in terms
of the asymptotic limiting densities.

The model we have introduced here allows for a rigorous analysis of the
{\em microscopic} structure of a shock of type 1 (with a discontinuity in 
the vacancy density), provided that some special 
conditions on the hopping parameters and shock densities are satisfied. 
For this shock one finds the same 
behaviour as for the ASEP, suggesting that also for nongeneric values
of the limiting densities of the shock the behaviour would be similar.
Shocks of type 2 (constant vacancy density) are microscopically unstable,
they correspond to diffusive spreading as in the symmetric exclusion
process.

Part of our investigation is devoted to the problem of steady state
selection in the open system \cite{Krug91a}. 
Unlike for driven diffusive system
with a single species of particles \cite{PTCP}
there is no developed theory of boundary-induced phase transitions
for systems with two or more conservation laws.
Using the known phase diagram of the ASEP and the microscopic
properties of the shock in our particular model
we have derived the stationary phase diagram.
So our model, even though degenerate, may serve as a testing ground
for any general theory. 
With a view on some of our earlier results 
on driven diffusive two-species systems
with open boundaries we note that we found that the
heuristic approach to steady-state selection of Ref. \cite{two-chanel}
reproduces the
independently derived phase diagram.

An intriguing observation is that
the $u$-density in the phase diagram depends continuously
on the left $v$-density throughout the maximal current phase. In order
to explain the significance of this observation we note that the phase
diagram of single-species systems can be predicted from the
current-density relation in terms of {\em boundary densities} \cite{Kol,min}, which in
general are unknown functions of the {\em boundary rates} of the
model. On the other hand any numerical study of a given model yields
the phase diagram in terms of the boundary rates, not the boundary
densities. Therefore a comparison between theoretical predictions and
numerical observations requires knowledge of the nonuniversal
relationship between boundary densities and boundary rates. In the
ASEP and generally in single-species systems the order parameter (the
bulk density) in
the maximal current phase does not depend on the left boundary
density. Hence it is not possible to measure the relationship between
the left boundary density and the boundary rates of the model. This in
turn makes it impossible to predict the phase diagram in terms of the
boundary rates. In the present two-species model, however, the left
boundary density can be measured in the maximal current phase as a
function of the system parameters through measuring $u$. 

This leads to
an unexpected offspin of our investigation: by regarding the
two-species model as an exclusion process with two kinds of vacancies
(by identifying vacancies with particles and the two kinds of
particles with vacancies).
Generalizing other single-species systems in a manner similar
to what is done here, viz., leaving their dynamics unchanged by
just introducing tagged diffusive vacancies, may provide a means of
measuring those postulated boundary densities in terms of the
model parameters even in parts of the phase diagram where the
order parameter itself does not depend on the boundary densities.

Finally we remark that in principle the processes $A\rightarrow B$ or 
$B\rightarrow A$ could also be excluded (i.e. $\gamma^{+(-)}_{l(r)}=0$)
from our process. We only keep them
for the sake of generality. If one wants to
choose the boundary conditions to mimic certain boundary densities these
rates are in general non-zero. However, in our model these processes
don't play an important role unlike in the so-called bridge model
where they are responsible for spontaneous symmetry breaking \cite{bridge}.     
The precise mechanism for this phenomenon and a quantitative description,
however, still remain a major challenge in the study of two-species
systems with open boundaries.

\begin{acknowledgments}
A. R\'akos thanks the Deutsche Forschungsgemeinschaft (DFG)
for financial support. We thank V. Popkov for useful discussions.
\end{acknowledgments}
\appendix

\section{Solution of the hydrodynamic equations with finite viscosity}
\label{visc}

The viscous hydrodynamic equations for the steady state coming from
the method described in Ref. \cite{two-chanel}
(see Sec. \ref{subsec:steady_selection}) are the
following: 
\begin{align}
\label{visc1}
(p-q)v'(2v-1) &= a\frac{p+q}{2} v'', \\
\label{visc2}
(p-q)(u'v+v'u)&=a\frac{p+q}{2} (u''v-v''u) + ac\left( (1-v)u''+uv'' \right).
\end{align}
Substituting $u=(1-v)\phi$ in (\ref{visc2}) and using (\ref{visc1}) we
arrive to the following equation for $\phi$: 
\begin{equation}
\label{eqphi}
v(1-v)\phi'=\frac{a(p+q)}{2(p-q)} \left( v(1-v)\phi'' - 2vv'\phi'
\right) + \frac{ac}{p-q}\left((1-v)^2\phi''-2(1-v)\phi'v'\right).
\end{equation} 
One can immediately see here that $\phi=\text{const.}$ is a solution
if it fits both boundaries, i.e., if
$\phi_\text{left}=\phi_\text{right}$. When this does not hold (this is
the case in general) then 
one has
\begin{equation}
\frac{d}{dx}\ln|\phi'|=\frac{2v'}{1-v}+\frac{v}{\frac{a}{p-q}
\left(\frac{p+q}{2}v+c(1-v)\right)}.  
\end{equation} 
Integrating from the right boundary we get
\begin{equation}
\ln|\phi'(y)|= \ln|\phi'(1)|
-\int_y^1\frac{v}{\frac{a}{p-q}\left(\frac{p+q}{2}v+c(1-v)\right)}dx -
2\int_{v(y)}^{v_\text{right}} \frac{dv}{1-v}. 
\end{equation} 
Assuming that the sign of $\phi'$ does not change in the system we arrive to
\begin{equation}
\label{phiv}
\phi'(y)=\phi'(1)\left( \frac{1-v_\text{right}}{1-v(y)}\right)^2
\exp\left(-\frac{p-q}{a}\int_y^1\frac{v}{\frac{p+q}{2}v+c(1-v)}dx\right). 
\end{equation} 
For strictly positive $v$ in the bulk of the system
one has $\phi'(y)\to 0$ as $a\to 0$ for any fixed $y<1$ thus the limiting $\phi$
profile is flat everywhere apart from the right boundary, which
implies  $\phi_\text{bulk}=\phi_\text{left}$.

There is no positive lower bound for $v$ if and only if
$v_\text{right}=0$ (dashed line in figure \ref{fig:phase_diagram}). In
this case $v(x)\to 0$ everywhere as $a\to 0$  
apart from the left boundary (provided $v_\text{left}\neq 1$) which
implies $\phi''(x)\to 0$ for any 
fixed $x>0$
according to (\ref{eqphi}). We still have to show that $\phi$ does not
have a discontinuity in the $a\to 0$ limit at the left boundary. For
this we evaluate $\phi'(0)$ and show that it does not
diverge. Similarly to (\ref{phiv}) one has
\begin{equation}
\phi'(0)=\phi'_\text{bulk}\left( \frac{1}{1-v_\text{left}}\right)^2
\exp\left(-\frac{p-q}{a}\int_0^y\frac{v}{\frac{p+q}{2}v+c(1-v)}dx\right),
\end{equation}
where the upper limit of the integration $0<y\leq 1$ is arbitrary
because of the fast decay of the integrand. Since the integrand is non-negative  the exponential cannot diverge which gives 
\begin{equation}
\lim_{a\to 0}\phi'(0)<\infty.
\end{equation}
This means that the $\phi$ profile is the linear function connecting
the two boundary values. The $u$ profile, however, has a discontinuity
at the left boundary according to (\ref{phi}) and jumps from
$u_\text{left}$ to $u_\text{left}/(1-v_\text{left})$. From here it
goes linearly to the right boundary value $u_\text{right}$.

\end{document}